\documentclass[preprint,showpacs,preprintnumbers,amsmath,amssymb]{revtex4}
\usepackage{graphicx}
\usepackage{dcolumn}
\usepackage{bm}
\usepackage{epsfig}

\begin{document}

\title[More on the Asymmetric Infinite Square Well]{More on the Asymmetric Infinite Square Well: \\Energy Eigenstates with Zero Curvature}

\author{L. P. Gilbert} \email{lagilbert@davidson.edu}
\affiliation{Physics Department, Davidson College, Davidson, NC
28035 USA}

\author{M. Belloni} \email{mabelloni@davidson.edu}
\affiliation{Physics Department, Davidson College, Davidson, NC
28035 USA}

\author{M. A. Doncheski} \email{mad10@psu.edu}
\affiliation{Department of Physics, The Pennsylvania State
University, Mont Alto, PA 17237 USA}

\author{R. W. Robinett} \email{rick@phys.psu.edu}
\affiliation{Department of Physics, The Pennsylvania State
University, University Park, PA 16802 USA}

\begin{abstract}
We extend the standard treatment of the asymmetric infinite square
well to include solutions that have zero curvature over part of
the well. This type of solution, both within the specific context
of the asymmetric infinite square well and within the broader
context of bound states of arbitrary piecewise-constant potential
energy functions, is not often discussed as part of quantum
mechanics texts at any level. We begin by outlining the general
mathematical condition in one-dimensional time-independent quantum
mechanics for a bound-state wave function to have zero curvature
over an extended region of space and still be a valid wave
function. We then briefly review the standard asymmetric infinite
square well solutions, focusing on zero-curvature solutions as
represented by energy eigenstates in position and momentum space.
\end{abstract}


\maketitle

\section{Introduction}
One of the most standard one-dimensional problems in quantum
mechanics is that of the infinite square well (ISW) given by the
potential energy function:
\begin{equation} V(x) = \left\{
\begin{array}{ll}
0 & \mbox{for $0<x<L$} \\
\infty & \mbox{otherwise}\;.
\end{array}
\right. \label{infinite_well_potential}
\end{equation}
The position-space wave functions of this system can be easily
shown to be: $\psi_n(x)=\sqrt{2/L}\sin(n\pi x/L)$, where $n=1, 2,
\ldots$ for $0<x<L$ and $\psi_n(x)=0$ otherwise. Despite this
model's ease of solvability, the ISW lacks some of the salient
features necessary to promote a general understanding of more
complicated systems in quantum mechanics. Since $V=0$ over the
entire width of the well, the classical value of the kinetic
energy, and hence the magnitude of the local value of the
classical momentum, does not vary with position. As a consequence,
the `wiggliness' (local value of the wave number) and amplitude of
the corresponding quantum-mechanical wave function must remain
constant over the entire width of the well. This model therefore
does not display the more general relationship between the kinetic
and potential energies as a system with a spatially-varying
potential energy.

The relatively simple extension to an asymmetric infinite square
well (AISW) augments the standard ISW model and begins to address
this issue by adding a constant potential energy `hump' over part
of the well:
\begin{equation}
      V(x) = \left\{ \begin{array}{ll}
+\infty & \mbox{for $x<-a$} \\
0  &  \mbox{for $-a<x<0$}\;\;\;\;\;\;\textrm{Region\;I}\\
+V_0  &  \mbox{for $0<x<+b$}\;\;\;\;\;\;\textrm{Region\;II} \\
+\infty  &  \mbox{for $+b<x$}
\end{array} \right.
\, . \label{eq:pot_aisw}
\end{equation}
The AISW, therefore, has an obvious spatial variation of the
potential energy, which accounts for change in the `wiggliness'
and amplitude of the wave function in the two regions. This
extension has obvious pedagogical benefits, and has been exploited
on a variety of levels
\cite{krane,morrison,robinett_book,harrison_book}. Besides being
of pedagogical interest, experiments exciting coherent charge
oscillations from just such asymmetric quantum well structures
have been reported \cite{coherent_charge}, which make use of both
standard types of solutions (both the $E>V_0$ and $E<V_0$
solutions) in the produced wave packet.

In a previous paper in this Journal \cite{robinett_asymmetric},
two of us considered the AISW by focusing on the comparison of the
classical and quantum-mechanical probability densities (both in
position and momentum space). Simple time-spent arguments for
$E>>V_0$ led to a geometrical probability of $a/(a+b)$ and
$b/(a+b)$ for Region I and Region II, respectively. In addition,
momentum-space results were discussed that yielded peaks that
qualitatively matched the classically-expected results.

In this paper, we describe an extension to the standard treatment
of the asymmetric infinite square well by considering wells in
which zero-curvature ($E-V_0=0$) wave functions naturally occur
for suitable choices of $a$, $b$, and $V_0$, and the quantum
number, $n$. While previous authors have used the zero-curvature
case to show that the unperturbed infinite square well cannot have
a zero-energy solution \cite{bowen,yinji}, cases in which
zero-curvature solutions are valid bound-state solutions have
seldom been considered \cite{zc_paper}, even though a linear
radial wave function, $u(r)$, naturally occurs in low-energy
S-wave scattering from finite potentials \cite{sakurai} used to
model low-energy nuclear scattering \cite{blatt_weisskopf}. In
Section 2 we briefly review solutions of the Schr\"{o}dinger
equation for regions of constant potential energy, including the
zero-curvature solution. We then describe in Sections 3 and 4 the
results for the asymmetric infinite square well, considering
separately the solutions for energies less than, greater than, and
equal to the potential energy step. We determine the probability
densities for the zero-curvature solutions in Section 5, and in
Section 6 we calculate and analyze the zero-curvature wave
functions in momentum space.

\section{Solutions to the Time-independent Schr\"{o}dinger Equation}
\label{schrodinger_section}

To begin the discussion of allowable solutions to the
time-independent Schr\"{o}dinger equation in one dimension,
consider a potential energy function that does not change with
position.  When this is the case, the potential energy function is
a constant in that region and the Schr\"{o}dinger equation
becomes:
\begin{equation}
\left[\,-\frac{\hbar^2}{2m}\frac{d^2}{dx^2}+V_0\,\right]\,\psi(x)=E\,\psi(x)\;,
\end{equation}
which can be written as
\begin{equation}
\left[\,\frac{d^2}{dx^2}+\frac{2m(E-V_0)}{\hbar^2}\,\right]\,
\psi(x)=0\;. \label{simple_tise}
\end{equation}

For our analysis, it is convenient to define
\begin{equation}
k\equiv\sqrt{2mE/\hbar^2}\;, \label{eq:k}
\end{equation}
\begin{equation}
\chi\equiv\sqrt{2m|V_0|/\hbar^2}\;, \label{eq:chi}
\end{equation}
\begin{equation} q\equiv\sqrt{2m(E-V_0)/\hbar^2}\;,
\label{eq:q}
\end{equation}
and
\begin{equation}
\kappa\equiv\sqrt{2m(V_0-E)/\hbar^2}\;. \label{eq:kappa}
\end{equation}

In Eq.~(\ref{simple_tise}) there are \textit{three} cases to be
considered: $E<V_0$, $E>V_0$, and $E=V_0$ (the zero-curvature
solution). In these three cases, the Schr\"{o}dinger equation and
its solution reduce to:
\begin{equation}
\left[\,\frac{d^2}{dx^2}-\kappa^2\,\right]\,
\psi(x)=0\;\;\;\;\;\;\;\rightarrow\;\;\;\;\;\;\psi(x)=Ae^{\kappa
x}+Be^{-\kappa x}\;, \label{eq:tunneling}
\end{equation}

\begin{eqnarray}
                     \left[\,\frac{d^2}{dx^2}+q^2\,\right]\,
\psi(x)=0\;\;\;\;\;\;\rightarrow\;\;\;\;\;\;& \psi(x)=A\cos(qx)+B\sin(qx)\nonumber\\
                     &\textrm{or}\;\;\nonumber\\
                     &\psi(x)=Ae^{iqx}+Be^{-iqx}\;,
\label{eq:oscillating}
\end{eqnarray}
and
\begin{equation}
\frac{d^2}{dx^2}\,
\psi(x)=0\;\;\;\;\;\rightarrow\;\;\;\;\psi(x)=Ax+B\;,
\label{eq:linear}
\end{equation}
for $E<V_0$, $E>V_0$, and $E=V_0$, respectively. When $E<V_0$, the
curvature of the wave function is such that
$\frac{1}{\psi(x)}\frac{d^2\psi(x)}{dx^2}>0$ and hence the wave
function curves away from the axis (positive curvature for
$\psi(x)>0$ and negative curvature for $\psi(x)<0$), and
exponentially decays, grows, or does both, depending on boundary
conditions. For $E>V_0$ the curvature of the wave function is such
that $\frac{1}{\psi(x)}\frac{d^2\psi(x)}{dx^2}<0$ and the wave
function is oscillatory (negative curvature for $\psi(x)>0$ and
positive curvature for $\psi(x)<0$). For $E=V_0$, which in general
is not considered except as an inflection point between regions of
positive and negative curvature, the curvature of the wave
function is zero, and the wave function is a constant or is
linear.

\section{The Asymmetric Infinite Square Well}
To illustrate the results of Section \ref{schrodinger_section}, we
focus on the asymmetric infinite square well (AISW) as defined by
the potential energy function defined in Eq.~(\ref{eq:pot_aisw}).
Beginning with the $E<V_0$ case, when we apply the boundary
conditions at $-a$ and $b$, we have the wave function in each
region:
\begin{equation}
\psi_{\textrm{\tiny{I}}}(x)=A\sin(k[x+a])\;, \label{eq:psiA4}
\end{equation}
and
\begin{equation}
\psi_{\textrm{\tiny{II}}}(x)=\frac{C}{2}\left[\,e^{\kappa[x-b]} -
e^{-\kappa[x-b]}\,\right]= C\sinh(\kappa[x-b])\;. \label{eq:psiB4}
\end{equation}
Matching these two pieces of the wave function at $x=0$,
$\psi_{\textrm{\tiny{I}}}(0)=\psi_{\textrm{\tiny{II}}}(0)$ and
$\psi^\prime_{\textrm{\tiny{I}}}(0)=\psi^\prime_{\textrm{\tiny{II}}}(0)$,
we find:
\begin{equation}
\kappa\tan(ka) = -k\tanh(\kappa b)\;. \label{eq:energy5}
\end{equation}
Eq.~(\ref{eq:energy5}) can then be used to determine the allowed
energy eigenstates since $k$ and $\kappa$ are both related to the
energy via Eqs.~(\ref{eq:k}) and (\ref{eq:kappa}). As an example
of a typical set of position-space wave functions, in Figure 1 we
have chosen $V_0=33$ and $\hbar=2m=1$ to visualize these states
(properly normalized).

For the $E>V_0$ case, we find that the wave function in each
region is
\begin{equation}
\psi_{\textrm{\tiny{I}}}(x)=A\sin(k[x+a])\;, \label{eq:psiA2}
\end{equation}
and
\begin{equation} \psi_{\textrm{\tiny{II}}}(x)=C\sin(q[x-b])\;,
\label{eq:psiB2}
\end{equation}
and again by matching these two pieces of the wave function at
$x=0$, we find the energy-eigenvalue condition:
\begin{equation}
q\tan(ka) = -k\tan(qb)\;. \label{eq:energy}
\end{equation}
For $E>V_0$, the wave functions of the AISW have some interesting
features as shown in Figure 1.  For the first five states,
corresponding to $E<V_0$, the wave function is mostly confined to
the left half of the well. As the energy increases above the step
height, as shown by the remaining states in Figure 1, the wave
function now extends over the entire well with a noticeable change
in `wiggliness' and amplitude across the middle of the well. As
was pointed out in \cite{robinett_asymmetric}, there are also
special cases (that occur for much higher energy states with these
parameters, and are not shown) in which there is an antinode at
$x=0$ which yields the same amplitude for the wave function across
the well.

The same states shown in Figure 1 in position space are shown in
Figure 2 in momentum space visualized using the momentum-space
probability density. For the states with $E<V_0$, we note two
peaks ($p=\pm\hbar k$ where $k$ is defined by Eq.~(\ref{eq:k})) in
the probability density that get farther apart as the energy, and
hence the momentum, increases. For the states with $E>V_0$, four
peaks ($p=\pm\hbar q$ and $p=\pm\hbar k$ where $q$ and $k$ are
defined by Eqs.~(\ref{eq:q}) and (\ref{eq:k}), respectively) arise
with the two peaks associated with the smaller momentum values
($p=\pm\hbar q$) having a noticeably larger probability than the
two peaks associated with the larger momentum values ($p=\pm\hbar
k$), as expected from classical time-spent arguments.

\section{The Zero-curvature Solution}
For $E=V_0$, the wave function in each region is
\begin{equation}
\psi_{\textrm{\tiny{I}}}(x)=A\sin(k[x+a])\;, \label{eq:psiA6}
\end{equation}
and
\begin{equation} \psi_{\textrm{\tiny{II}}}(x)=C[x-b]\;,
\label{eq:psiB6}
\end{equation}
and by matching these two pieces of the wave function at $x=0$, we
find:
\begin{equation}
A\sin(ka)=-C b\;, \label{eq:match5}
\end{equation}
and
\begin{equation}
kA\cos(ka)=C\;. \label{eq:match6}
\end{equation}
By dividing Eq.~(\ref{eq:match5}) by Eq.~(\ref{eq:match6}) and
simplifying, we have the energy-eigenvalue condition:
\begin{equation}
\tan(ka) = -kb\;, \label{eq:energy6}
\end{equation}
and since $E=V_0$, Eq.~(\ref{eq:energy6}) is also a condition on
$V_0$:
$\tan(\sqrt{{\scriptstyle{2mV_0/\hbar^2}}}\,a)=-\sqrt{{\scriptstyle{2mV_0/\hbar^2}}}\,b$\;.
Zero-curvature solutions are, therefore, only possible for special
AISWs with suitable values of $V_0$, $a$, and $b$, \textit{and}
the quantum number, $n$. It is easy to conclude that for an AISW,
there exists either one or (more likely) no zero-curvature states
in the energy spectrum.  For example, in an AISW with $a=b=3$ and
$\hbar=2m=1$, we find that zero-curvature states occur with: $V_0
= 0.457318\;(n = 1),\;2.682149\;(n = 2),\;7.073234\;(n = 3),
13.654351\;(n = 4),\;22.427917\;(n = 5),\;33.394444\;(n =
6),\;46.55409\;(n = 7),\;61.906917\;(n = 8)$, \;79.452954\;(n =
9).  The allowed zero-curvature wave functions for $n=1$ through
$n=9$ with $a=b=3$ and the special $V_0$ values given above are
shown in Figure 3.

If we consider the wave functions in Region II with the limiting
cases in which $\kappa\rightarrow 0$ and $q\rightarrow 0$, we find
that,
$\psi_{{\textrm{\tiny{II}}}\,{\scriptscriptstyle{(E<V_0)}}}(x)=
\psi_{{\textrm{\tiny{II}}}\,{\scriptscriptstyle{(E>V_0)}}}(x)
\approx\psi_{{\textrm{\tiny{II}}}\,{\scriptscriptstyle{(E=V_0)}}}(x)$.
However, we must be careful not to assume this means that
$\lim_{\kappa\rightarrow
0}\,\psi_{{\textrm{\tiny{II}}}\,{\scriptscriptstyle{(E<V_0)}}}(x)$
and $\lim_{q\rightarrow
0}\,\psi_{{\textrm{\tiny{II}}}\,{\scriptscriptstyle{(E>V_0)}}}(x)$
\textit{always} give valid solutions to the $E=V_0$ case. They do
not. Only in the cases where we have \textit{already tuned} the
potential energy well to yield a wave function energy of $E=V_0$
does taking the limit of the $E>V_0$ and $E<V_0$ wave functions
yield the exact zero-curvature solution.

\section{Position-space Probability Distributions}

While Eq.~(\ref{eq:psiA6}) and Eq.~(\ref{eq:psiB6}) give the wave
function in each region, they must still match at $x=0$ and
therefore we use Eq.~(\ref{eq:match5}) to yield
\begin{equation} \psi(x)
= \left\{ \begin{array}{ll}
A\sin[(k(x+a)] & \mbox{for $-a\leq x \leq 0$} \\
-A\frac{\sin(ka)}{b}(x-b) & \mbox{for $0 \leq x \leq +b$}
\end{array}
\right. \label{eq:match_position_psi_zc}
\end{equation}
for $\psi_{\textrm{\tiny{I}}}(x)$ and
$\psi_{\textrm{\tiny{II}}}(x)$, apart from normalization. To
normalize we require that
\begin{equation}
\!\!\!\!\!\!\!\!\!\!\!\!\!\!\!\!\!\!\!\!\!\!\!\! |\psi(x)|^2\,dx =
A^2\int_{-a}^{0} \sin^2[(k(x+a)]\,dx +
\left(\frac{A\sin(ka)}{b}\right)^2\int_{0}^{+b} (x-b)^2\,dx=1\;,
\end{equation}
and we therefore find that
\begin{equation}
A=\left[\,\frac{a}{2}\left(1-\frac{\sin(2ka)}{2ka}\right) +
\frac{b\sin^2(ka)}{3}\,\right]^{-1/2}\;. \label{eq:norm}
\end{equation}

Shown in Figure 4 are the normalized probability densities for the
zero-curvature states ($n=1$ through $n=9$) for the same nine
wells used in Figure 3.

We also find by direct integration that the distribution of
probability in the two regions is
\begin{equation}
\!\!\!\!\!\!\!\!\!\!\!\!\!\!\!\!\!\!\!\!\!\!\!P_{{\textrm{\tiny{I}}}}=
\frac{a\left(1-\sin(2ka)/2ka\right)}{2[\,\frac{a}{2}(1-\sin(2ka)/2ka)
+b\sin^2(ka)/3\,]}=A^2\left[\frac{a}{2}\left(1-\frac{\sin(2ka)}{2ka}\right)\right]\;,
\label{prob_left}
\end{equation}
and
\begin{equation}
\!\!\!\!\!P_{\textrm{\tiny{II}}}=
\frac{b\,\sin^2(ka)}{3[\,\frac{a}{2}(1-\sin(2ka)/2ka)
+b\sin^2(ka)/3\,]}=A^2\left[\frac{b\,\sin^2(ka)}{3}\right]\;.
\label{prob_zc}
\end{equation}
We can analyze these results by noticing that as $n$ increases for
the zero-curvature solution, Eq.~(\ref{eq:energy6}) implies that
$k\rightarrow\frac{(2n-1)\pi}{2a}$; this in turn yields
$\sin(ka)\approx \pm 1$ (and therefore $\sin^2(ka)\approx 1$) and
$\sin(2ka)\approx 0$. Eq.~(\ref{prob_left}) and
Eq.~(\ref{prob_zc}) reduce to
$P_{\textrm{\tiny{II}}}\approx\frac{3a}{(3a+2b)}$ and
$P_{\textrm{\tiny{II}}}\approx\frac{3b}{(3a+2b)}$, respectively,
for large $n$. In the case where $a=b$, such as in all of the
Figures, these equations yield $P_{{\textrm{\tiny{I}}}}\approx
0.6$ and $P_{\textrm{\tiny{II}}}\approx 0.4$. This approximation
is already accurate to within 0.5\% for the $n=6$ zero-curvature
state and is borne out by Figure 4.

\section{Momentum-space Probability Distributions}
The zero-curvature wave functions in momentum space are
straightforward to calculate from the position-space wave
functions via the Fourier transformation:
\begin{equation}
\phi(p)=\frac{1}{\sqrt{2\pi \hbar}} \int_{-a}^{+b}
\psi(x)\,e^{-ipx/\hbar}\,dx\;,
\end{equation}
and by using the wave function in
Eq.~(\ref{eq:match_position_psi_zc}), we have
\begin{eqnarray}
\phi(p)=&&\frac{A}{2i\sqrt{2\pi \hbar}}
\,\int_{-a}^{0}\,\left(\,e^{ik(x+a)}-e^{-ik(x+a)}\,\right)
\,e^{-ipx/\hbar}\;dx \nonumber\\
&&- \frac{A\sin(ka)}{b\sqrt{2\pi\hbar}}
\,\int_{0}^{+b}\,(x-b)\,e^{-ipx/\hbar}\,dx\;.
\end{eqnarray}
We can define these terms such that $\phi(p) \equiv\phi_{(+)}(p) +
\phi_{(-)}(p) + \phi_{(0)}(p)$ where the $\phi_{(\pm)}(p)$ wave
functions come from the $\exp[i(\mp k -p/\hbar)x]$ terms
integrated over the left side of the well, and the $\phi_{(0)}(p)$
wave function comes from the linear term integrated over the right
side of the well. The explicit expressions are given by
\begin{equation}
\!\!\!\!\!\phi_{(\pm)}(p) = \pm \frac{A}{\sqrt{2\pi
\hbar}}\left(\frac{ai}{2}\right) e^{\mp ika}e^{+i(p\pm \hbar
k)a/2\hbar}\left[ \frac{\sin[(p\pm \hbar k)a/2\hbar]}{(p\pm \hbar
k)a/2\hbar} \right]\;,
\end{equation}
 and
\begin{equation}
\phi_{(0)}(p) = \frac{A}{\sqrt{2\pi
\hbar}}\left(\frac{\hbar^2\sin(ka)}{bp^2}\right) \left[1-ipb/\hbar
- e^{-ipb/\hbar}\right]\;.
\end{equation}
From these forms, it is clear that as the quantity $ka$ increases,
the momentum-space wave functions become more noticeably peaked at
$p = \pm \hbar k$ and $p=0$.

In Figure 5, we visualize the probability density in momentum
space for the zero-curvature states ($n=1$ through $n=9$) for the
same nine wells used in Figure 3.

Using these results, it is also easy to show that
\begin{equation}
\int_{-\infty}^{+\infty} |\phi_{(+)}(p)|^2\,dp =
\int_{-\infty}^{+\infty} |\phi_{(-)}(p)|^2\,dp=
A^2\left(\frac{a}{4}\right) \;,
\end{equation}
\begin{equation}
\int_{-\infty}^{+\infty} |\phi_{(0)}(p)|^2\,dp =
A^2\left(\frac{b\sin^2(ka)}{3}\right)\;, \label{p_zero_squared}
\end{equation}
\begin{equation}
\!\!\!\!\!\!\!\!\!\!\!\int_{-\infty}^{+\infty} \left[
\phi_{(-)}^{*}(p)\, \phi_{(+)}(p) + \phi_{(-)}(p)\,
\phi_{(+)}^{*}(p) \right]\,dp =
-A^2\left(\frac{a\sin(2ka)}{4ka}\right)\;,
\end{equation}
and
\begin{equation}
\!\!\!\!\!\!\!\!\!\!\!\!\!\!\!\! \int_{-\infty}^{+\infty} \left[
\phi_{(0)}^{*}(p)\, \phi_{(+)}(p) + \phi_{(+)}^{*}(p)\,
\phi_{(0)}(p)+\phi_{(0)}^{*}(p)\, \phi_{(-)}(p) +
\phi_{(-)}^{*}(p)\, \phi_{(0)}(p)\right]\,dp = 0\;,
\label{eq:integral_0_pm}
\end{equation}
where $A$ was previously defined in Eq.~(\ref{eq:norm}).

We first note that the position of the maxima of the central
`peak' alternates between a single peak at $p=0$ and two peaks at
$p=\pm\alpha$ for odd and even values of $n$, respectively. For
$p\approx 0$, the numerical value of the cross terms in the
momentum-space probability density, the integrand in
Eq.~(\ref{eq:integral_0_pm}), alternates in sign as a function of
$n$, thereby alternating constructive and destructive interference
with the direct term, $|\phi_{(0)}(p)|^2$, as a function of $n$.
The effect of these cross terms in the momentum-space probability
density is therefore evident in the alternating position of the
maximum of the central `peak,' despite the fact that these terms
do not yield any overall contribution to the probability as shown
by Eq.~(\ref{eq:integral_0_pm}).

Also note that the amount of probability in the various `peaks' is
directly comparable to the probabilities of being in the left or
right regions of the well. We can easily make these assignments by
noticing that the momentum-space probability of measuring the
particle with $p\approx 0$ is $A^2b\sin^2(ka)/3$, which
corresponds to the probability of being localized in the right
side of the well since the probabilities calculated in
Eq.~(\ref{prob_zc}) and Eq.~(\ref{p_zero_squared}) are identical.
We can similarly compare the momentum-space probability for the
$p\approx \pm\hbar k$ peaks to the probability of the particle
being localized in the left side of the well, again finding that
these are identical.

Given our well parameters, for large $n$ we find a momentum
distribution that approaches 30\% for the left peak (corresponding
to a negative momentum in the left side of well), 40\% for the
central peak (corresponding to zero momentum in right side of
well), and 30\% for the right peak (corresponding to a positive
momentum in the left side of well), again agreeing with the
60\%-40\% split in the position-space probability for the left and
right regions of the well.

\section{Conclusion}

We have outlined the conditions in which an asymmetric infinite
square well, and other piecewise-constant potential wells, can
have a valid zero-curvature bound-state wave function. These
position-space wave functions, and their momentum-space
counterparts, are easily determined and visualized. They have
readily calculable probability relationships which allow a direct
comparison between the position- and momentum-space probabilities
for each region of the well. Other infinite square well
extensions, such as infinite well plus Dirac delta function cases
\cite{zc_paper}, can also exhibit zero-curvature solutions for the
right well parameters.  A more general analysis of these cases is
straightforward, and will be addressed elsewhere
\cite{other_paper}.

Zero-curvature states can also serve an important pedagogical
purpose as a way to easily extend the standard treatment of the
AISW and other piecewise-constant potential wells.  These special
cases help further elucidate the connection between the potential
energy function, the quantized energy eigenvalue, and the
resulting form of the wave function in one-dimensional
quantum-mechanical systems. While zero-curvature solutions are an
intuitively natural interpolation between the much more frequently
discussed oscillatory and tunneling solutions, the unfamiliar
mathematical form of the one-dimensional Schr\"{o}dinger equation
in Eq.~(\ref{eq:linear}) for these parameters does catch many
students by surprise.

A recent (unintentional) educational experiment, where first-year
graduate students in physics were asked to consider just such a
problem, showed that less than 20\% could successfully find the
linear solution in Eq.~(\ref{eq:linear}) and the resulting
energy-eigenvalue condition in Eq.~(\ref{eq:energy6}) on an exam.
The most common mistake among these students, as is often seen in
such physics education trials, was an attempt to fit `standard'
solutions (in this case those in Eq.~(\ref{eq:tunneling}) and
Eq.~(\ref{eq:oscillating})) to this slightly different situation.
Having as wide an array of example problems allows instructors to
better probe and shape student understanding of quantum-mechanical
bound-state problems \cite{rick_qmvi}.

In addition, the procedures we have outlined can be used to
determine specifications for quantum wells, like those used in
Ref.~\cite{coherent_charge}, so zero-curvature wave functions can
be experimentally observed and their unique properties exploited.
Such zero-curvature states are of interest because they may
enhance scattering of wave packets in these special asymmetric
infinite square wells \cite{sprung}. This is yet another example
of the experimental realization of `designer' quantum wave
functions \cite{meekhof}.

\begin{acknowledgments}
We would like to thank Wolfgang Christian for useful conversations
regarding this work. LPG and MB were supported in part by a Research
Corporation Cottrell College Science Award (CC5470) and MB was also
supported by the National Science Foundation (DUE-0126439).
\end{acknowledgments}

\newpage
\begin{figure}[h]
\epsfig{file=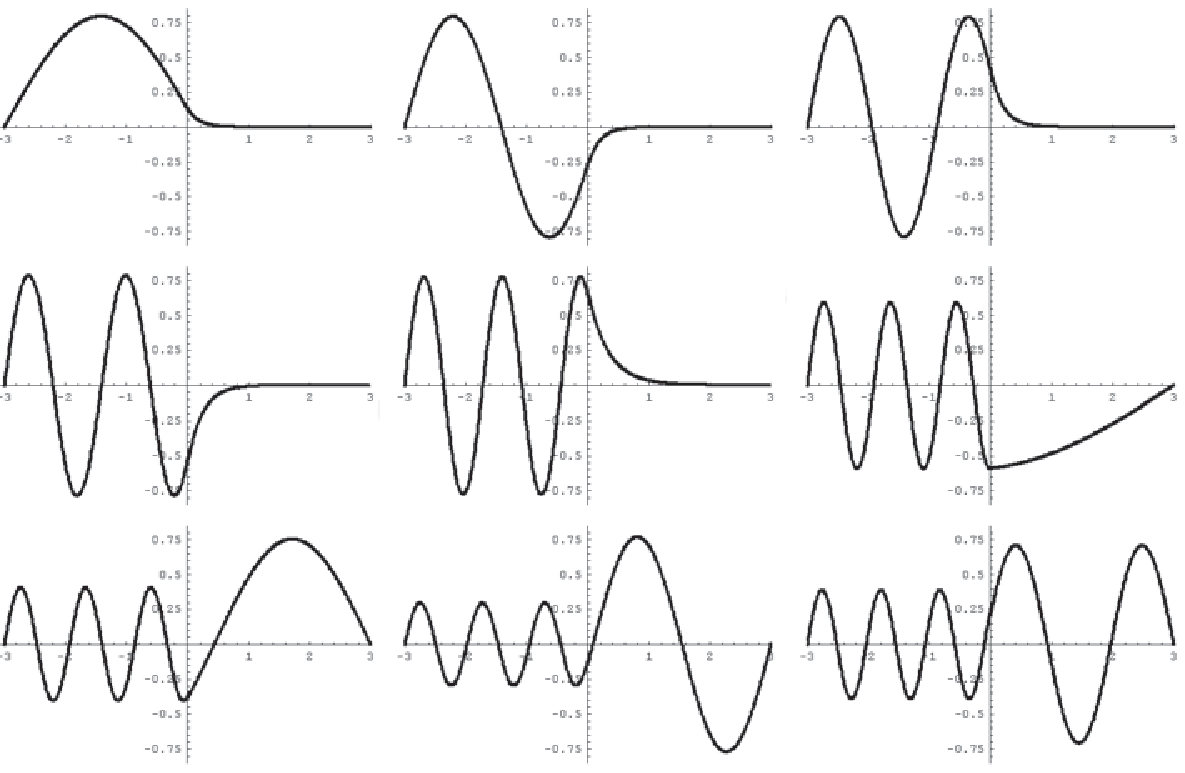,width=16cm,angle=0}

\caption{The first nine wave functions for the asymmetric infinite
square well. In all of the images $\hbar=2m=1$, $V_0=33$, and
$a=b=3$.}
\end{figure}

\newpage
\begin{figure}[h]
\epsfig{file=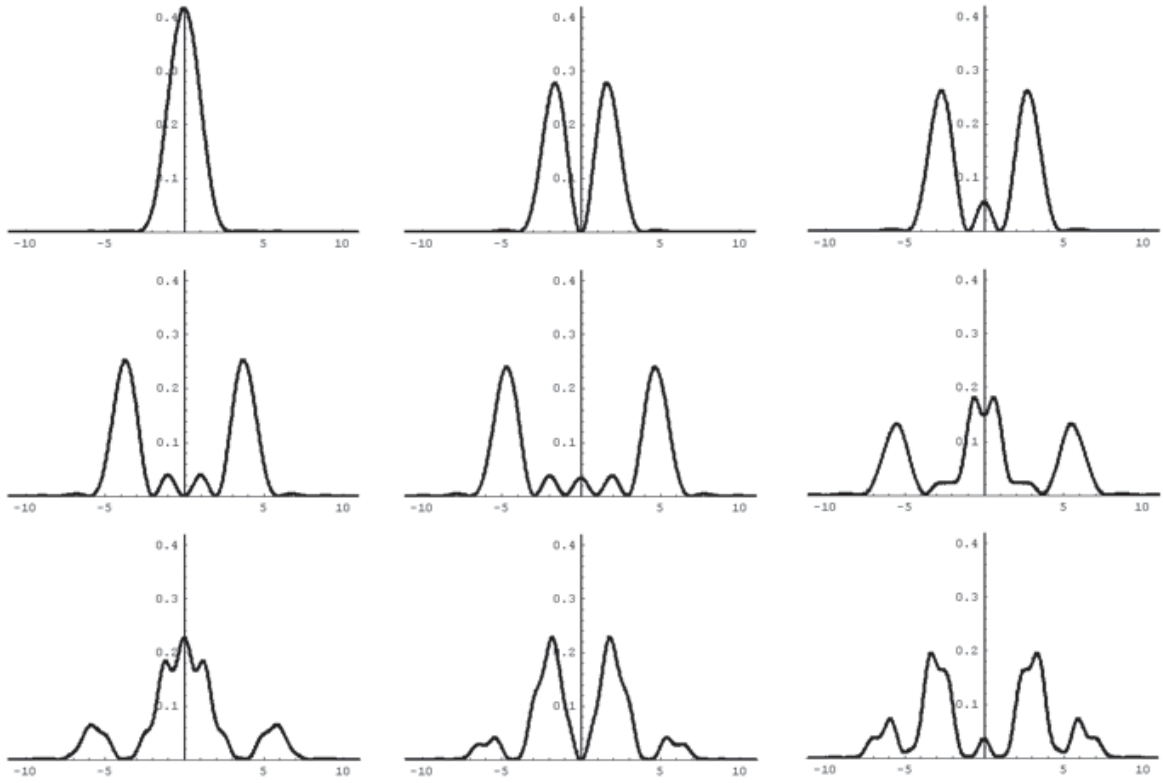,width=16cm,angle=0}

\caption{The first nine probability densities in momentum space for
the asymmetric infinite square well corresponding to the same
parameters as in Figure 1.}
\end{figure}

\newpage
\begin{figure}[h]
\epsfig{file=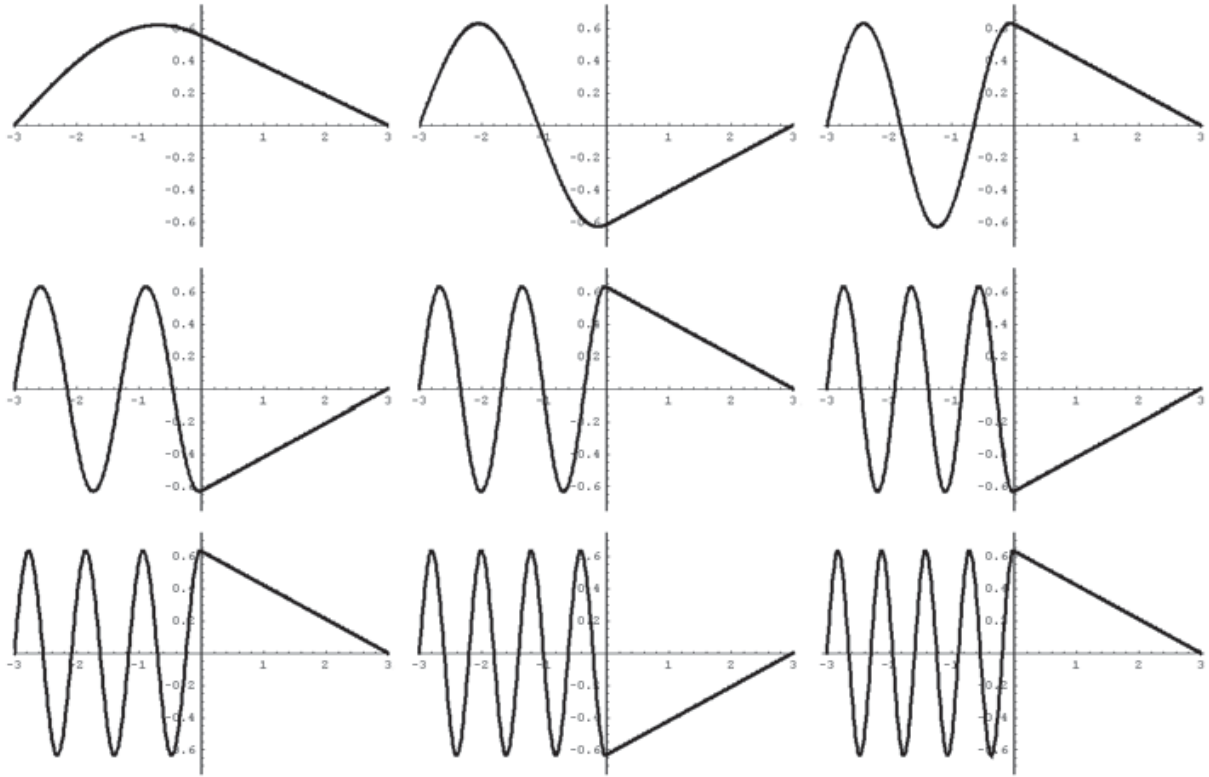,width=16cm,angle=0}

\caption{Nine zero-curvature wave functions with quantum numbers
$n=1$ through $n=9$ for asymmetric infinite square wells with
$a=b=3$ and different values of $V_0$ that satisfy
Eq.~(\ref{eq:energy6}). In all of the images $\hbar=2m=1$.}
\end{figure}

\newpage
\begin{figure}[h]
\epsfig{file=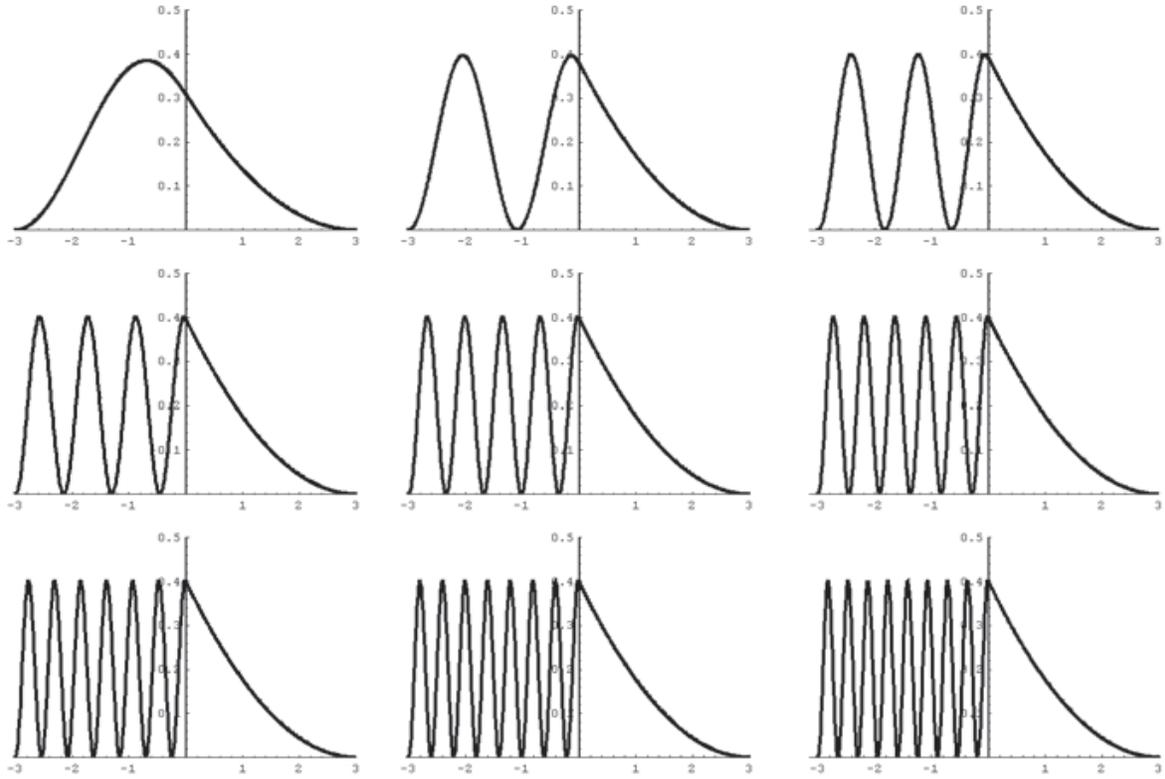,width=16cm,angle=0}

\caption{The probability densities for zero-curvature wave functions
($n=1$ through $n=9$) for the same nine wells used in Figure 3.}
\end{figure}

\newpage
\begin{figure}[h]
\epsfig{file=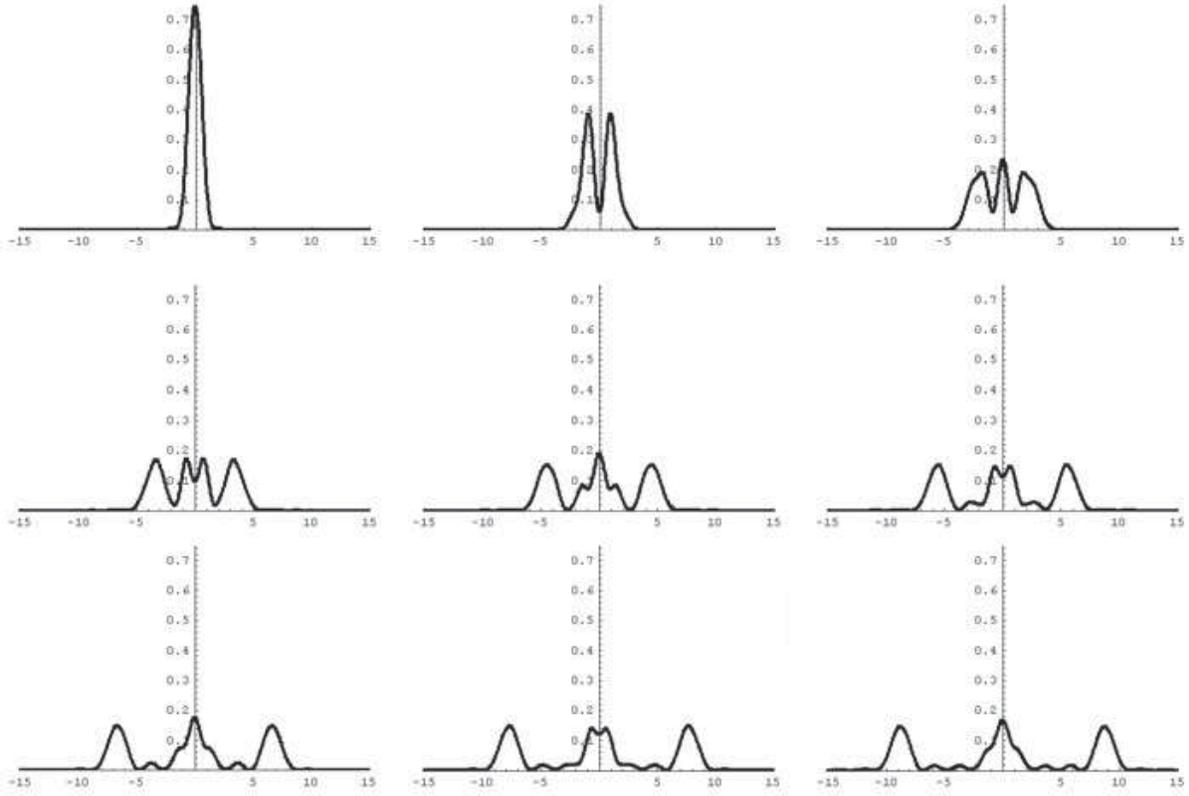,width=16cm,angle=0}

\caption{The zero-curvature probability densities in momentum space
($n=1$ through $n=9$) for the same nine wells used in Figure 3.}
\end{figure}

\end{document}